# Characterization of Metastatic Tumor Formation by the Colony Size Distribution[1]

## Steve P. Lee,[2] Ji-Rong Sun, Hong Qian, William H. McBride, and H. Rodney Withers

*Department of Radiation Oncology, University of California, Los Angeles, California 90095[S. P. L., J.-R. S., W. H. M., H. R. W.] and Department of Applied Mathematics, University of Washington, Seattle, Washington 98195[H. Q.]*

## ABSTRACT

**Knowledge regarding the kinetics of metastatic tumor formation, as related to the growth of the primary tumor, represents a fundamental issue in cancer biology. Using an *in vivo* mammalian model, we show here that one can obtain useful information from the frequency distribution of the sizes of metastatic colonies in distant organs after serial sectioning and image reconstruction. To explain the experimental findings, we constructed a biophysical model based on the respective growth patterns of the primary tumor and metastases and a stochastic process of metastatic colony formation. Heterogeneous distributions of various biological parameters were considered. We found that the elementary assumption of exponential forms of growth for the primary tumor and metastatic colonies predicts a linear relation on a log-log plot of a metastatic colony size distribution, which was consistent with the experimental results. Furthermore, the slope of the curve signifies the ratio of growth rates of the primary and the metastases. Non-exponential (Gompertzian and logistic) tumor growth patterns were also incorporated into the theory to explain possible deviation from the log-log linear relation. The observed metastasis-free probability also supported the assumption of a time-dependent Poisson process. With this approach, we determined the mechanistic parameters governing the process of metastatogenesis in the lungs for two murine tumor cell lines (KHT and MCaK). Since biological parameters specified in the model could be obtained in the laboratory, a workable metastatic "assay" may be established for various malignancies and in turn contribute in formulating rational treatment regimens for subclinical metastases.**

## INTRODUCTION

Metastatic formation in a cancer patient represents a detrimental event despite the advancement of localized therapy for the primary disease. Since modern diagnostic technology still leaves much to be desired for the detection of metastases, the oncologists often effect a "prophylactic" treatment strategy for the presumably undetected metastases (defined as *micrometastases*, or *subclinical metastases*). So far, the clinical practice has been based mainly upon empirical observations (1). More recently, statistical analysis of clinical data has implicated the usefulness of low-dose radiation therapy previously thought to be ineffective for the control of subclinical disease (2). In order to provide a rational basis in designing treatment regimens for subclinical metastases, we aim to construct a mechanistically oriented quantitative theory describing the metastasis process, and test its validity with *in vivo* mammalian experimentation. We wish to modify our theoretical model in a controlled fashion by adding complexity one step at a time. The ultimate aim is to develop a plausible mathematical model that reflects reality as much as possible, yet complexity should be minimized so that it is practical. The particular model could be rendered implausible easily by contradictory data, but the methodology should remain incorruptible.

Therefore, with regard to the metastatic problem, we hypothesize that, in general, the relative sizes of metastatic colonies in a distant organ reflect the temporal order of their initial arrivals, while the size-specific numerical frequencies correlate directly with the primary tumor bulk at the time of the metastatic "shedding". In such way, the information leading to the kinetic relationship between the primary tumor and its metastatic progenies is "archived" as the metastatic colony size distribution, dictated by their respective growth behaviors and the probability of the metastatic events.

## MATERIALS AND METHODS

### *In Vivo* Murine Experiment

**Animal Preparation and Tumor Cell Lines.** C3H female mice from the specific pathogen free animal colony in the Experimental Division of the UCLA Department of Radiation Oncology were used in the *in vivo* experiments. When necessary, the animals were anesthetized via intraperitoneal injection of Nembutal (62.5 mg/kg). They were maintained in cages and in accordance with the UCLA animal protection guidelines. Two different murine tumor cell lines, KHT (sarcoma) and MCaK (breast cancer), were used.

**Ascertain Growth Behavior of Primary Tumors.** Each mouse was inoculated with a fixed amount of tumor cells subcutaneously at its flank. The sizes of the primary tumors were monitored serially in time by measuring the length, width, and depth of each palpable tumor with micrometer, and multiplying all three to obtain the volume. The "average diameter" was arbitrarily set to be the cube root of this volume - we elected not to account for the shape of the tumor and likewise assumed that the volume calculated is related directly to the number of clonogenic cells, and ignored the possible existence of a necrotic portion in which the cells were most likely dead. The graph of the primary tumor volume was plotted against time after the first measurable nodule occurred (around 3-mm average diameter), since the time from tumor cell inoculation to the first appearance of palpable tumor was seen to vary widely.

---

[1] Supported in part by American Cancer Society Physician's Research Training Award (to S. P. L.).

[2] To whom requests for reprints should be addressed, at Department of Radiation Oncology, University of California, Los Angeles, 200 UCLA Medical Plaza, Suite B265, Los Angeles, CA 90095-6951. Phone: (310) 825-0128; Fax: (310) 794-9795; E-mail: stevelee@radonc.ucla.edu.

**Obtaining Metastatic-free Probability Curve.** After primary tumor implantation, the sizes of the primary tumors were measured and the average diameter obtained as described above. All mice underwent surgical resection of their primary tumor after the tumor grew to a pre-set size, randomly assigned to different categories of 6, 7, 8, 9, 10, 11, 12, 14, 16 cm average diameter. However, the actual size of each tumor upon resection was measured, and the final averaged value was obtained for that size category (bin) when calculating the frequency of metastasis for that particular bin. The animals were then allowed to live either to imminent death or up to 100 days post implantation, at which time they would be sacrificed (except those assigned to the experiment described in the next paragraph). Only mice with no sign of local recurrence of its primary tumor were included in the subsequent analysis. Upon their death, the lungs of these animals were surgically excised and fixed in Bouin's solution, which enabled direct visualization of metastatic colonies under low-power magnification. If no tumor colony was visible at all, the animal was considered operationally to be free of any metastasis at the time of the primary tumor excision. A MFP[3] curve as a function of the size of primary tumor, $N$, was thus generated.

**Obtaining Metastatic Colony Size Distribution.** In this experiment, mice underwent resection of their primary tumor after it grew to a designated size, then sacrificed at pre-set number of days after the surgery; some were allowed to keep the primary tumor *in situ* until spontaneous death or sacrifice. Only mice with no sign of local primary tumor recurrence were included in the subsequent analysis. Upon death, the lungs of all animals were surgically excised and fixed in formalin, paraffin-embedded and stained with hematoxylin and eosin. Serial slide sections of 5-μm thickness were then obtained and examined meticulously under high-power microscope. Contiguous cross-sections of individual metastatic colonies were traced and their areas obtained via digital imaging technique, described as follows: A video camera (*Javelin CCD* camera, Javelin Electronics, Torrance, CA) is attached to a microscope equipped with multiple power fields. Each slide was scanned in a raster motion manually, and all visible metastatic colonies (up to the highest power field) were identified. The video images were captured by a frame grabber (*Oculux TCX*, Coreco Inc., St-Laurent, Quebec), and the digitized data sent to a computer. Using image analysis software (*Optimas* software, BioScan, Inc., Edmonds, WA) that allows one to quantify the area of an enclosed region traced on the digitized image, the areas of the cross sections of all colonies were obtained for each slide. These results were submitted to the computer and stored in a data base spreadsheet. Volumes were then computed by "stacking" the images for each metastatic colony in contiguous sections, summing the corresponding cross-section areas, and multiplying the slide thickness. In such fashion, a histogram of the volume distribution of metastatic colonies was obtained.

## Mathematical Model

**General Approach.** After formulating a hypothesis based on biophysical argument, the resulting mathematical expression is solved analytically to obtain a closed-form solution specifying the quantitative relationship between the variable in question and various biophysical parameters. If an analytical solution cannot be obtained, numerical simulation is performed (*Mathcad 8*, MathSoft, Inc., Cambridge, MA). Fittings of data with the theoretical formulas are done via non-linear regression algorithm (*STATISTICA* software, StatSoft, Inc., Tulsa, OK).

**Metastatic-free Probability.** We begin with the most elementary assumption that both the primary tumor and the metastatic colonies grow exponentially, with parameters $\lambda$ and $\mu$ characterizing the respective growth rates. Next, we focus on quantifying the probability of the *eventual establishment* of the metastatic colonies, since studies using intravital videomicroscopy has suggested that the rate-limiting step in the post-vascular phase of the "metastatic cascade" to be at the final growth stage in the destined organs (3). Furthermore, the endpoints, which bypass the entire chain of events after the cancer cells detach from the primary tumors until they start to grow in the distant organs (effectively treating such multi-steps as a "black box"), are of primary clinical significance after all. Since the process of metastatic colony formation is known to be very inefficient (4), we assume such probability to be very small, although at any moment it is in direct proportion with the primary tumor size. Finally, we expect the numbers of these events occurring in disjoint time intervals to be independent. Mathematically, these assumption are equivalent to stating that the probability of establishing a metastatic colony follows a *nonhomogeneous Poisson process* (5) with an "intensity function" in proportion to $N(t)$, the primary tumor size. The proportional constant, termed $\eta$, thus represents the rate of metastatic colony formation, or *metastatic rate* for short, and has a dimension of *number of colonies formed per time per primary tumor cell / size*. It is another biological parameter defined precisely in this model and corresponds to the crudely described "metastatic potential" in clinical oncology. Hence, we find[4]

$$MFP(t) = \exp\{-(\eta/\lambda)[N(t) - N(t_0)]\}, \quad (1)$$

where $t_0$ is the starting time of the Poisson process. Eq. (1) predicts MFP to follow a sigmoid curve when plotted as a function of the logarithm of $N$ (*i.e.*, time $t$, assuming exponential primary tumor growth: $N(t) = N_0 e^{\lambda t}$, where $N_0$ is the size of the primary at $t = 0$). On a plot of ln*MFP* vs. $N$, a linear function is expected with a slope dictated by the ratio $\eta/\lambda$. However, if we incorporate heterogeneity into $\eta$ by assuming it to be normally distributed among a population cohort with a mean, $\eta_0$, and variance, $\sigma_\eta^2$, it can be found that[4]

$$\ln MFP(t) \sim -(\eta_0/\lambda)[N(t) - N(t_0)] + (\sigma_\eta^2/2\lambda^2)[N(t) - N(t_0)]^2 \quad (2)$$

**Metastatic Colony Size Distribution.** Fig. 1A shows the rather heuristic argument why the theoretical metastatic colony size distribution curve can be a straight line on a log-log plot, if both the primary tumor and all the colonies grow exponentially at constant rates. We wish to explore the biological meaning of this curve, i.e., expressing its slope in terms of the biological parameters specified in the mechanistic model. Define $R(m,t)$ as the number of distinct metastatic colonies *per size m* at time $t$ (thus called "*size density function*", the integral of which over the *m*-axis gives the total number of colonies at time $t$). If each metastatic colony grows with a constant rate of $\mu$, we can formulate a partial differential equation[4]:

$$\partial R(m,t)/\partial t = -\partial[\mu m R(m,t)]/\partial m .$$

With appropriate initial conditions, the solution is given as

$$R(m,t) = (\eta/\mu)N_0 e^{\lambda t} m^\kappa , \quad (3)$$

where $\kappa = -[(\lambda/\mu)+1)]$ is the slope of the straight line on a ln$R$ vs. ln$m$ plot. If the heterogeneity in $\mu$ is considered with a mean value of $\mu_0$, a straight line also results on a log-log plot, with the slope governed by $-[(\lambda/\mu_0)+1)]$ as a leading-order approximation. This is derived in detail by assuming $\mu$ to be either normally distributed among all metastatic colonies or subject to random fluctuations in time, with $\mu_0$ as the *mean* metastatic growth rate. In both scenarios, we again find a linear relationship (to a first-order approximation, at least) on a log-log plot, with the slope $\kappa$ as a function of $\lambda$ and $\mu_0$.[4] It is feasible to obtain the solutions for the complicated theoretical equations dealing with heterogeneous $\mu$ whether numerically or via analytical calculation. Biologically meaningful parameters (i.e. the slope) can still be inferred readily. Finally, analytical solutions for $R(m,t)$ are also found separately using Gompertzian and logistic growth models for the primary tumor.[4]

---

[3] The abbreviations used are: MFP, metastasis-free probability; MFSP, metastasis-free survival probability; MCP, metastasis control probability.

[4] S. P. Lee, H. R. Withers, and H. Qian, manuscript in preparation

In practice, when we obtain laboratory or clinical data, it is usually the *number* of colonies within a particular range of size that we possess. Such "*size frequency distribution*", $F(m,t)$, can be derived from the above-mentioned size-specific "density", $R(m,t)$, to show:

$$F(m,t) \equiv mR(m,t) = (\eta/\mu)N_0 e^{\lambda t} m^{-\lambda/\mu}. \quad (4)$$

That is, when plotted on $\ln F$ vs. $\ln m$, the slope of $\ln F(m,t)$ differs from that of $\ln R(m,t)$ by -1. When plotted on a semi-log graph against $\ln m$, it should reflect an *exponential distribution* (5), with the parameter $\lambda/\mu$.

When the primary tumor is excised successfully, one would expect the subsequent metastatic colony size distribution to be altered in time as established metastases grow but no new ones are created (Fig. 1B). Using the most elementary assumption of a uniform growth rate, $\mu$, for all colonies, we can figure out the evolutionary course of $R(m,t)$ in time. Specifically, if we define $t_{Ex}$ as the time the primary tumor is excised, then, at time $t$ ($t > t_{Ex}$), we would expect the *smallest* metastatic colonies to result from the growth of metastatic cells which begin at time $t_{Ex}$, thus having a size (defined as $m_a(t)$, a function of $t$) described by $\ln m_a(t) = \mu(t - t_{Ex})$, where we assume that all colonies begin to grow as a single cell (i.e. $m_a(t_{Ex}) \equiv 1$). Furthermore, the *number* of these smallest colonies is seen to be preserved in time, while each continues to grow exponentially at a uniform rate. However, if growth rates among metastatic colonies are not uniform, one can expect the persistent presence of colonies with size smaller than $m_a(t)$.

## RESULTS

**Primary Tumor Growth Behavior.** In a pilot experiment, $2 \times 10^5$ MCaK cells were inoculated in 119 mice, and the growth curve was subsequently obtained. Fig. 2A shows the growth curve, with the primary tumor size as a function of time after the first measurable tumor appears. The growth resembles exponential pattern for the early and major portion of its history (dark symbols). The solid curve comes from fitting this portion of the observed data for the early growth with the theoretical prediction based on the exponential model. The averaged growth rate was found to be approximately 0.25 day$^{-1}$, or a tumor doubling time of 2.8 days. Fig. 2B shows the growth curves for a second experiment with 150 mice, each implanted with $3 \times 10^5$ MCaK cells, with averaged growth rate at the early, exponential phase found to be approximately 0.18 day$^{-1}$, or a tumor doubling time of 3.9 days. Apparently, both sets of data showed significant deviation from the exponential pattern toward the later phase of the tumor growth (open symbols), consistent with growth retardation. This demonstrates the inadequacy of the exponential model to describe the *entire* history of the tumor growth. In fact, fitting with logistic and Gompertzian functions for the entire set of data points were done, and more complicated biological parameters were obtained accordingly based on our theoretical models.[4]

**Metastatic-free Probability Curve.** For the 150 mice implanted with $3 \times 10^5$ MCaK cells, all were randomly assigned to undergo surgical resection of their primary tumor at pre-set sizes. After excluding those with local recurrences and other causes (premature death of unclear cause, loss of identification tag on the animal's ear, etc.), a total of 106 animals were eligible for this part of the experiment. Fig. 3A shows the *MFP* as a function of the average primary tumor volume actually measured within each assigned size-category (with the number of animals in each category varied from 9 to 18), but on a logarithmic scale (thus corresponding to time-axis linearly). A monotonically decreasing sigmoid curve is seen with the inflection point occurring at a primary tumor size of 870 mm$^3$. Fig. 3B is generated based on the same data, but presented as $\ln MFP$ vs. $N$, the primary tumor volume. One can see that the data appear to fit better a linear-quadratic function, which is consistent with a theory based on a normally distributed $\eta$ (Eq. (2)). Recall that the primary tumor growth rate ($\lambda$) for this cohort of mice was found to be 0.18 day$^{-1}$ (Fig. 2B). From Fig. 3B, we obtained $\eta_0/\lambda = 1.15 \times 10^{-3}$ mm$^{-3}$. Thus, the mean metastatic rate ($\eta_0$) can be calculated to be $2.07 \times 10^{-4}$ d$^{-1}$mm$^{-3}$, and the standard deviation, $\sigma_{\eta}$, for the presumed Gaussian heterogeneity in $\eta$ is $9.3 \times 10^{-5}$ d$^{-1}$mm$^{-3}$. If one assumes that a tumor of 1 cm$^3$ contains about $10^9$ cells, then $\eta_0$ becomes $2.07 \times 10^{-10}$ per day per cell. That is, for MCaK tumor, this translates to approximately an *average* of 1.15 metastatic events per $10^9$ cells (1 cm$^3$) of primary tumor. This does not mean that there is 1.15 metastatic events in *every* 30 doublings (since $2^{30} \sim 10^9$) of the primary tumor. It merely means that, for a large number of animals each with a primary tumor of $10^9$ cells, while some might have no metastasis at all, others would have experienced 1, 2, 3 ... in an essentially infinite number of such events. Overall, there is *on the average* 1.15 metastatic events which have occurred for each animal. One can see that such notion as "metastatic potential" is quite crude, and thus we do not advocate its use. In contrast, the "metastatic rate", $\eta$, as defined operationally in our model, possesses a clear, mechanistic, meaning, and perhaps should be employed routinely for the quantification of the metastatic tendency of a particular malignancy.

**Metastatic Colony Size Distribution.** A subset of mice transplanted with $2 \times 10^5$ KHT or MCaK tumor cells were sacrificed with primary tumor in place (*in situ*), and their lungs harvested for the purpose of constructing the metastatic colony size distribution curve. Those designated to undergo excision of their primary tumors at approximately 10 mm average diameter were sacrificed at 0, 1, 4, 6, 8, 12 days post-resection of their primary tumors. As the procedure involved in obtaining the colony size distribution for each animal was quite tedious, and the purpose of the study was for validation of principle, data based on only a few mice was obtained. Here we present the result for a mouse injected with KHT cells without primary tumor excision, and two mice injected with MCaK cells with primary tumor excision at 10 mm average diameter and sacrificed at 6 days and 12 days post-surgery, respectively.

For the mouse with KHT inoculation and without primary tumor excision, a total of 277 metastatic colonies were found,

with size ranging from minimum of 1.76 x $10^{-6}$ mm$^3$ to maximum of 3.47 mm$^3$. Fig. 4A shows the histogram display of the number of colonies distributed among 21 bin categories plotted on a natural logarithmic scale. When the data are presented in such a semi-log fashion, they appear to follow the theoretical curve based on the prediction of Eq. (4), i.e. an exponential distribution (solid curve). A parameter of 0.16 was found. The dashed curve corresponds to an exponential distribution extrapolated retroactively after fitting a log-log plot with a linear function (log$F(m,t)$ in Fig. 4B), with the ratio $\lambda/\mu$ found to be 0.25, consistent with micrometastases growing 4 times faster than the primary.

To follow closely how much the data points deviate from the theoretical prediction, the residuals for ln$R(m,t)$ are plotted in Fig 5A. We see that deviation occurs predominantly at both extremes of the small and the large size categories. Figs. 5B and 5C show the results of fitting of the size density function with Gompertzian and logistic models for primary tumor growth, respectively. Indeed, better fitting of the data points can be achieved with these kinds of non-exponential primary tumor growth, with the expected more complex form of biological parameters specified in these models (each with one more parameter than the exponential model).

Fig. 6 shows the result of the two mice implanted with MCaK cells, which then underwent primary tumor excision at a size of 10.2 mm and 10.5 mm average diameter, respectively. The first mouse was allowed to live another 6 days, and the second 12 days, until they were sacrificed. The growth curve for the individual primary tumor, from the time it became palpable until it was excised, was obtained in both cases, and the growth rates were found to be approximately identical: $\lambda = 0.27$ d$^{-1}$. One can see from Fig. 6 that their individual metastatic colony size distribution appears to contain two phases. The solid symbols in both curves represent the linear size distribution of the larger metastatic colonies and reflect the kinetics of metastatogenesis prior to the excision of the primary tumor, similar to the schematic drawing of Fig. 1B. We may use this portion of the curve to extrapolate the metastatic growth rate, $\mu$, for either mouse. For the first, it is calculated to be 0.66 d$^{-1}$, and for the second, 0.29 d$^{-1}$ (corresponding respectively to tumor doubling time of 1.0 d and 2.4 d). Even though there were 6 more days of waiting for the second mouse to be sacrificed, its slower growth rate apparently offset such difference, since the expected positions on the abscissa of the log-log plot for the left-most end-points of the truncated ln$R(m,t)$ lines (numerically equals to $\ln m_a(t) = \mu(t - t_{Ex})$; see also Fig. 1B) turned out to be similar for the two mice (3.96 and 3.48, respectively, equivalent to the smallest metastases having grown to contain about 52 cells and 32 cells). In fact, if we adhere to our assumption that 1 cm$^3$ of tumor represents an aggregate of $10^9$ cells, then a single cell would be about $10^{-6}$ mm$^3$, of which the logarithm would be -13.8. The values of $m_a$ for the two mice determined above would thus be 5.2 x $10^{-5}$ mm$^3$ and 3.2 x $10^{-5}$ mm$^3$ (the logarithm would be -9.9 and -10.3, respectively). We can see from Fig. 3.15 that there was indeed an absence of smaller colonies in either mouse, ranging from single cell to about 45 cells (or 4.5 x $10^{-5}$ mm$^3$, of which the logarithm is -10), which agrees somewhat with the values of $m_a$ estimated above.

The rough calculations done so far, and likewise the "hand-waving" arguments used in Figs. 1A and 1B, pertain to the ideal assumption that all metastatic colonies within the same mouse grow at a uniform rate. Our experimental data can be interpreted to show otherwise, *i.e.* that the growth rates of the colonies are heterogeneously distributed and the value of $\mu$ found is at best an average, since clearly "truncated" straight lines for the size distributions were not obtained for the two animals that underwent primary tumor resection. Some colonies of smaller sizes (open symbols in Fig. 6) apparently got "left behind", failing to grow bigger as predicted based on a uniform $\mu$.

## DISCUSSION

**Primary Tumor Growth Behavior.** In our murine experiments, there was evidence that growth retardation of the primary tumors occurred mainly at a relatively *late* stage. Biological reasons for such behavior include lack of nutrition or oxygen, or some kind of negative feedback control in operation. When the entire set of data points for both experiments were used for fitting with theoretical growth models, a satisfactory result could be obtained for either logistic or Gompertzian model (with the former showing the best outcome). They might even warrant a more complicated description of the whole growth process, such as a "Gomp-ex" pattern advocated by Wheldon (6), or a two-phase stochastic model analyzed by Zheng (7). However, one must note that there were significant statistical and experimental errors for tumors of larger dimension (open symbols in Figs. 2A and 2B), which were not only few in number but each also contained massive central necrosis. Furthermore, fitting with non-exponential functions in general requires more parameters than that for an exponential form, which would introduce more degrees of arbitrariness into our theoretical model. This defies our purpose of maintaining as simple as possible our mechanistically oriented model, according to Occam's Razor. Most importantly, the primary purpose of our experimentation is to investigate the kinetics of *subclinical* metastatic formations, which occur mostly over the *early* major portion of the total growth duration of the primary tumor. We have shown in our animal experiment that such early primary growth can be modeled fairly well with the exponential form, from which the numerical value of $\lambda$ can be extracted. For MCaK tumors, for example, the averaged growth rate for the initial 20 days or so after palpable tumors first appear was found to be approximately 0.18 to 0.25 day$^{-1}$, or a tumor doubling time of 2.8 to 3.9 days. The possible effect of heterogeneity in $\lambda$, as modeled using non-negative Gaussian distribution, did not seem significant after numerical simulations using a wide range of variance.

**Metastatic-free Probability Curve.** Ubiquitous in nature, Poisson process has been used by several investigators in explaining the process of metastasis, albeit at different stages in the cascade and with various biological assumptions (8-12).

Our formulation predicts the probability of remaining free of any metastasis, *MFP*, to follow a sigmoid curve when plotted as a function of the logarithm of primary tumor size, $N$ (Eq. (1)). This was supported by the experimental data (Fig. 3A). The tumor size (around 870 mm$^3$ for MCaK cells) at which the *MFP* decreases sharply (i.e. the inflection point of the curve) will be governed by both primary tumor growth rate, $\lambda$, and metastatic rate, $\eta$, which may actually depend on time as well as vary among different individuals in a population. Indeed, when plotting a curve of ln*MFP* vs. $N$ for the case of constant $\eta$, a *linear* function is expected. Our experimental observation, however, agrees better with the prediction when $\eta$ is assumed to be normally distributed among the population cohort (Eq. (2), Fig. 3B), a reasonable biological assumption. In such a case, an additional *quadratic* effect is found due to the variance of $\eta$, so that ln*MFP* actually has a slower rate of decline as $N$ increases. By numerical simulation, the overall effect of heterogeneous $\eta$ on a *MFP* vs. ln$N$ plot can be shown to make the sigmoid curve a bit "flatter".

That a graph of *MFP* vs. ln$N$ should give rise to a sigmoid curve as predicted by our model may not be apparent when one examines clinical data. For example, Thames et al. (12) and Koscielny et al. (13-15) all presented statistical analyses of human breast cancer data and provided some illustrations of metastatic probability as a function of primary tumor size. The three-parameter maximum-likelihood fitting of the data by the former group and the log-probit representation by the latter have yielded useful information. In this paper we are concerned about the probability of no metastasis *at all* (*MFP*), as governed by our assumption of a Poisson process of metastatic colony establishment. Our experimental result was similar to the laboratory observation by Liotta et al. (8, Figure 5).

We have also obtained reasonable agreement between our approach and the statistical analysis of Koscielny et al. (13) for a set of *MFP* data. These investigators presented their analysis of 2684 breast cancer patients at Institut Gustave Roussy and gave a set of clinical data pertaining to *MFP*, which were extracted from a collection of metastasis-probability *vs.* time curves for different primary tumor sizes at initial treatment (Figure 1 in their paper). By extrapolating the eventual leveling-off levels of these curves back to the time of the initial treatment, a table was generated for "proportions of initiated metastases" (one minus which would be *MFP*) as a function of the tumor size (Table I in their paper). In this table, two sets of data were given: one from actuarial curves, the other from an assumption of a "lognormal" model. We fitted the part obtained from the actuarial curves with our theoretical expression of Eq. (2), and found that the ratio of the mean of metastatic rate, $\eta_0$, to the primary tumor growth rate, $\lambda$, to be about 0.018 cm$^{-3}$. Since these authors claimed the median tumor doubling time to be about 7 months (Table I in Ref. 14) - thus corresponding to $\lambda$ of about 0.1 mon$^{-1}$ - the mean metastatic rate turned out to be about 0.0018 mon$^{-1}$ cm$^{-3}$ per our analysis. Furthermore, from this ratio and Eq. (2) (and noting that $N(t_0)$ is relatively negligible), we could determine the tumor size at which *MFP* decreases to 50%, $N(t_{0.5})$, which turned out to be 38.5 cm$^3$. This value is comparable in magnitude with what the authors called the median "threshold" volume of 23.6 ml (95% confidence range from 0.14 to 4000 ml), after their log-probit analysis.

**Metastatic Colony Size Distribution.** We have shown the feasibility of extracting the growth rate of metastases, $\mu$, from size-distribution of the metastatic colonies (given as the frequency distribution, $F(m,t)$, or the size density function, $R(m,t)$). In the case of constant $\mu$, and assuming exponential growth for both primary tumor and the metastatic colonies, Eq. (3) is obtained, and predicts for a straight line on a graph of ln$R(m,t)$ vs. ln$m$. More importantly, the slope of this line can tell us something about the biological parameters set forth in our quantitative model: it reflects the ratio of the growth rates of the primary tumor and the metastases. This is one of the true merits of our "mechanistically" oriented model.

That a straight line is often an artifact introduced by a representation using log-log plot is *not* a valid criticism of our model here (it would be if all we do were to fit a straight line statistically upon the data, without caring about its biological meaning). Note that the frequency distribution depicted as the number of colonies counted over a specific bin size (thus given as $F(m,t)$, or $mR(m,t)$) gives rise to an exponential distribution when plotted against ln$m$ (Eq. (4)). The same biological information can thus be obtained from a semi-log plot of $F(m,t)$ vs. ln$m$ (Fig. 4A, *cf.* Fig. 4B).

Hence, the value of the metastatic growth rate, $\mu$, which is largely unknown whether in laboratory or clinical settings, can be inferred using this model if the primary tumor growth rate, $\lambda$, is known. The more realistic case of non-uniform metastatic growth rate has also been considered, and we again find a linear relationship (to a first-order approximation) on a log-log plot, with the slope dictated by the ratio $\lambda/\mu_0$. Thus, from Fig. 5A, this ratio for KHT cells was found to be 0.24 (i.e. an average growth rate of micrometastases which is about four times faster than the growth rate of the primary tumor), yet the phenomenon of a heterogeneous $\mu$ was not conspicuous in such an experiment.

On the other hand, we also excised the primary tumors at certain sizes from a subset of mice and removed their lungs at different time periods post-excision. Their individual metastatic colony size distribution appeared to contain two phases, as exemplified in Fig. 6 for MCaK cells. The solid symbols represent the linear size distribution of the larger metastatic colonies and reflect the kinetics of metastatogenesis prior to primary tumor excision. The open symbols in Fig. 6 correspond to smaller colonies, which may have resulted from their slower growth rates because, in all cases they must have been established before the excision, thus implying indirectly that the metastatic growth rates are not uniform.

Nevertheless, non-uniform metastatic growth rates might not be the only plausible explanation for Fig. 6. Note that the values of $m_a$ for the two mice as determined fall very close to the theoretical prediction based on an uniform $\mu$ for either animal. Furthermore, both values correspond to the size of the smallest colonies at the left-end of the open symbols, instead of the solid ones from which the straight lines are extrapolated. In either case, therefore, the deviation of the open symbols from the straight line may be a reflection of *shortage in numbers* of these smaller colonies rather than a consequence of their *slower growth*. This shortage in numbers cannot be attributed solely to our inability to detect and score the smaller metastatic colonies because, as already shown in our earlier experiment using the mouse without primary tumor excision, we are quite capable of identifying some extremely small colonies with our image analysis technique. We may have missed some, but not likely all of them. Thus, all metastatic colonies within the same animal could have grown at a relatively uniform rate, but some mechanism (perhaps as a consequence of the primary tumor excision) might have existed to exert an inhibitory effect on relatively smaller colonies preferentially.

When the graph of residuals (*i.e.* the difference of each data point from the theoretical prediction) of our metastatic colony size distribution for the KHT tumor is plotted (Fig. 4A), it appears that there is a *systematic* deviation from the straight line towards the logarithmically smaller sizes. Although there may be some underscoring of these minute micrometastases, the deviation is consistent with the fact that the theoretical curve is based on exponential tumor growth, whereas the primary tumor may instead approach a significant plateau during its late growth stage. While this may also explain partially the pattern of the open symbols in Fig. 6 for the case of MCaK primary tumors which were excised off the two animals, we note that their size at excision (about 1000 mm$^3$) corresponded well within the early exponential phase of the growth, as can be inferred from Fig. 2. At any rate, while we also have analytical solutions for the colony size distribution for non-exponential growth patterns of the primary tumor, there are other possible explanations for deviation from a straight line at smaller sizes, including: 1. Some biological mechanisms exist to eliminate the small metastases preferentially, 2. Some smaller metastases fail to grow further because they have an intrinsic growth rate limitation (which is insensitive to the consideration of a normally distributed $\mu$), 3. Experimental errors (small metastatic colonies are simply harder to detect), and 4. Secondary metastases from pre-existing metastases (which may in fact increase the number of the smaller metastases, but is extremely unlikely given the metastatic inefficiency characteristic of the primary tumor). These factors may not be totally independent, and any experimental attempt to determine their individual contribution must control well the possible confounding variables. When a specific biological factor is considered relevant, our mathematical model can be modified accordingly. This may necessitate more complicated analytical formulations, but can be quite feasible using modern computers for numerical results.

When taking into consideration non-exponential primary tumor growth by using the Gompertzian and the logistic models, we also obtain the analytical expressions for the colony size distribution, which depend explicitly on the mechanistic parameters in question. Both appeared to describe the laboratory data well (Figs. 5B and 5C). Kinetic parameters governing the metastatic process can thus be determined from the colony size distributions. In this way, the animal model could potentially be utilized as an "assay" method to quantify the metastatic potential of different malignant cell lines.

The quantitative model developed may also be applied in the clinical setting if one obtains the "volume" distribution of metastases from diagnostic imaging studies of cancer patients. Iwata et al. (16) have published a study with the metastasis size distributions obtained from clinical tomographic images, taken sequentially in time for a patient with a hepatocellular carcinoma and subsequent multiple intra-hepatic metastases. These investigators apparently considered the original primary tumor to be an entity distinct from the subsequent appearances of multiple tumors within the same organ (liver), and treated these lesions as its bona fide metastatic progenies. With such proviso, we have tried to explain their data with our own model and extract the biological parameters as specified. Their data were fitted with our model based on Gompertzian (perhaps more appropriate for these relatively large, clinically-detected lesions) as well as exponential growth patterns for the metastases. Biological parameters were obtained for both. In the analysis by Iwata et al. themselves, Gompertzian growth model for the tumors (both the primary and the metastases) was used, although discussions regarding exponential and power laws of growth were also given (16). However, they considered the primary tumor and its metastatic progenies to have identical growth rates and behavior. This was perhaps appropriate for the particular clinical example they tried to model, namely, with both primary and metastases being in the same organ. They further included additional complexities empirically, such as a fractal dimension of the tumor size (thus shedding of metastases from only the surface of the primary tumor could be considered), and secondary metastases from metastases themselves. In contrast, our model allows one to distinguish the probably existing differences between the growth behaviors (both the pattern and its associated parameters, e.g. the rate of growth) of the primary tumor and its metastases. For example, using the exponential model, we found $\lambda/\mu$, from their size-distribution data on post-diagnosis day 432, 559, and 632, to be 0.57, 0.60, and 0.68, respectively. That is, the ratio between the primary to average metastatic growth rates was about 2/3, not unity. So far, we have declined adding fractal dimension or other modifications of our elementary model to accommodate possible tumor structural or physiological factors. Likewise, secondary metastases are ignored. We would add these complexities beyond our elementary model one at a time, when more data from well-controlled experiments are available and warrant such endeavors.

In addition to these main differences between the approach of Iwata, et al. and ours, we also need to emphasize that here we aim to assess the kinetic relationship between primary tumor growth and *subclinical* metastatic formations. Thus, the exponential growth model alone may be adequate for describing the early-growth phase of the primary tumor and the metastases. Armed with such an elementary theoretical construct, we only need to know about relatively few biological parameters (i.e. $\eta$, $\lambda$, and $\mu$), based on which we can now elaborate and attempt to model the apparently more complicated, clinically obtained information, as exemplified in the subsequent chapters.

Since the task of obtaining detailed metastatic colony size distribution in our experiment is extremely tedious and labor-intensive, we have begun to develop an algorithm based on stereological principles (17-19) to expedite obtaining the size distributions in a more efficient manner by sampling two-dimensional histologic sections of organs involved by metastases. The clinical implication of cross-sectional sampling of the metastatic colony size distribution is quite obvious, in view of the fact that tomographic (i.e. 2-D cross-sectional) imaging is ubiquitous in medical diagnostic arena today. On the other hand, as computer technology improves, a complete set of data pertaining to the 3-D volume distribution of metastatic colonies may soon be easy to obtain readily in clinical or laboratory settings, thus obviating the need for sampling.

**Clinical Applications.** We have applied our quantitative model constructed so far in the clinical setting. For example, we formulated a theoretical expression for the metastasis-free survival probability (*MFSP*), defined as the probability of *detecting* no metastasis using a particular diagnostic technique that has a certain limiting resolution, as a function of time after the diagnosis of the presumably localized primary tumor. With the hypothesized Poisson process and its associated theory of waiting-time distribution, we have constructed a model correlating clinically observed *MFSP* data with various biological parameters. It is feasible to correlate clinical prognostic factors (age, sex, race, *etc.*) or pathologic findings (histologic features, hormonal status, molecular diagnostics entities, *etc.*) with our hypothesized biological parameters governing the process of metastatogenesis. In such a way, systematic analysis of metastatic behavior for human malignant tumors can complement the laboratory assay method mentioned above.

Another example we have explored was using the biological parameters specified in our mathematical model to determine the metastases dose-response relationship for a cytotoxic agent like radiation therapy. It is well-established that the *tumor control probability* (*TCP*) is related directly to the radiation dosage and the *tumor cell burden* (*i.e.* total number of cells in a tumor), and presents as a sigmoid curve when plotted against radiation dosage (20) - a consequence of random cell killing. Nevertheless, the relationship between the *metastasis control probability* (*MCP*) and the dose of prophylactic radiation treatment to an anatomical region at risk of subclinical metastases has generally remained elusive due to the lack of knowledge of the *metastatic cell burden*, $M$. Using our model, we can determine $M$ directly from $R(m,t)$ and express it as a function of the primary tumor size, $N$, and the biological parameters, $\eta$, $\lambda$, and $\mu$. The resulting plot of *MCP vs.* dose is subsequently found to be a sigmoid curve as well, although its steep portion could be "flattened" significantly due to the heterogeneity in $\mu$ or $\eta$ for a patient population. Thus, the clinical implication of a spread in these biological parameters is that relatively small radiation dosage may actually be quite beneficial for the treatment of subclinical metastases, in contrast to an apparent threshold effect introduced by assuming no heterogeneity. The flatter curve from assuming such heterogeneity agrees with the conclusion of Withers *et al.* (2) from statistical analysis of clinical data, and provides a rational basis for achieving some, albeit reduced, benefit from radiation doses less than those traditionally preferred for treating subclinical diseases.

With the advancement in modern molecular biology research, our approach can also be utilized to investigate in a systemic and quantitative manner the effects of any biological factor hypothesized to control the metastatic cascade. Equipped with our mechanistic model for metastasis formation, one could investigate some hypothetical cases - at least in theory - encountered in laboratory or clinical settings. For example, the model may allow investigators to analyze what they might consider as distinct biological mechanisms in the phenomenon of metastasis: one governing the *initiation* of the metastatic process ($t_0$), the second, the *rate* of such occurrence ($\eta$), the third, the behavior of the *source* of the metastases ($\lambda$), and finally, the growth behavior of the metastases themselves ($\mu$). When applying mechanistically oriented model like ours to real biological data, one must be careful in the interpretation of the results. Nevertheless, the model allows biologists to think logically in a systematic and controllable fashion. The interplay between the data fitting and the incorporation, modification, or elimination of mechanistic parameters within the logical framework so established, should constitute a worthwhile exercise in the study of the biology of metastasis.

In summary, we have developed a laboratory animal model to ascertain the metastatic behaviors of malignant tumors by obtaining the metastasis-free probabilities and the colony size distributions. A mechanistically oriented quantitative theory has been formulated and appeared to explain the experimental and clinical observations well. Such mechanistic modeling is gaining momentum in biomedical research (21), and oncology is no exception (22). The advantage of our approach (rather than an empirically derived statistical model based on curve-fitting) is the feasibility of obtaining various biological parameters governing the process of metastasis. By using the proposed model, different tumor types may be systematically assayed to analyze their specific metastatic potency, and in turn may allow appropriate therapeutic strategy to be planned in the clinical setting.

## ACKNOWLEDGEMENTS

We thank Julian Cook for discussion and assistance in developing the mathematical model.

# Figure Legends

Fig. 1. Schematic drawing showing that the theoretical metastatic colony size distribution curve to be a straight line on a log-log plot, if both the primary tumor and all the colonies grow exponentially at constant rates. Assume the primary tumor to grow exponentially to 10 times its original size per equal time interval $\Delta t = t_{i+1} - t_i$, during which each and every metastatic colony grows 2 times its original size. Furthermore, assume the metastatic efficiency to be one colony establishment per 100 primary tumor cells. Then, in $A$, the distribution of the number of colonies, depicted as $F(m,t)$, should be a straight line on a log$F$ vs. log$m$ display, with the slope governed by the growth rates of the primary tumor and the metastatic colonies. In time, this line should "travel" progressively. In $B$, the primary tumor is excised at time $t_5$. The straight line becomes "truncated" as it travels in time towards the right. The left upper "end-point" of each line stays at a constant level of $10^3$, representing the number of the smallest colonies, while each colony within this group enlarges in size from $2^0$ at time $t_5$ to $2^4$ at time $t_9$.

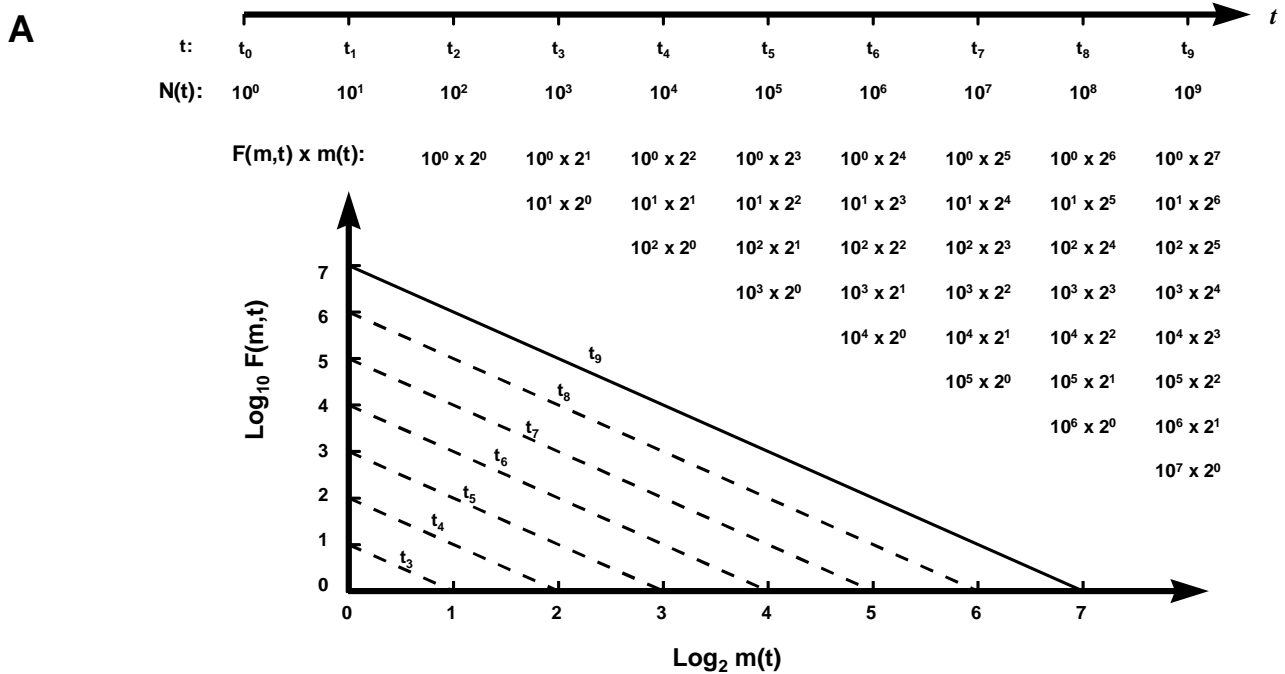

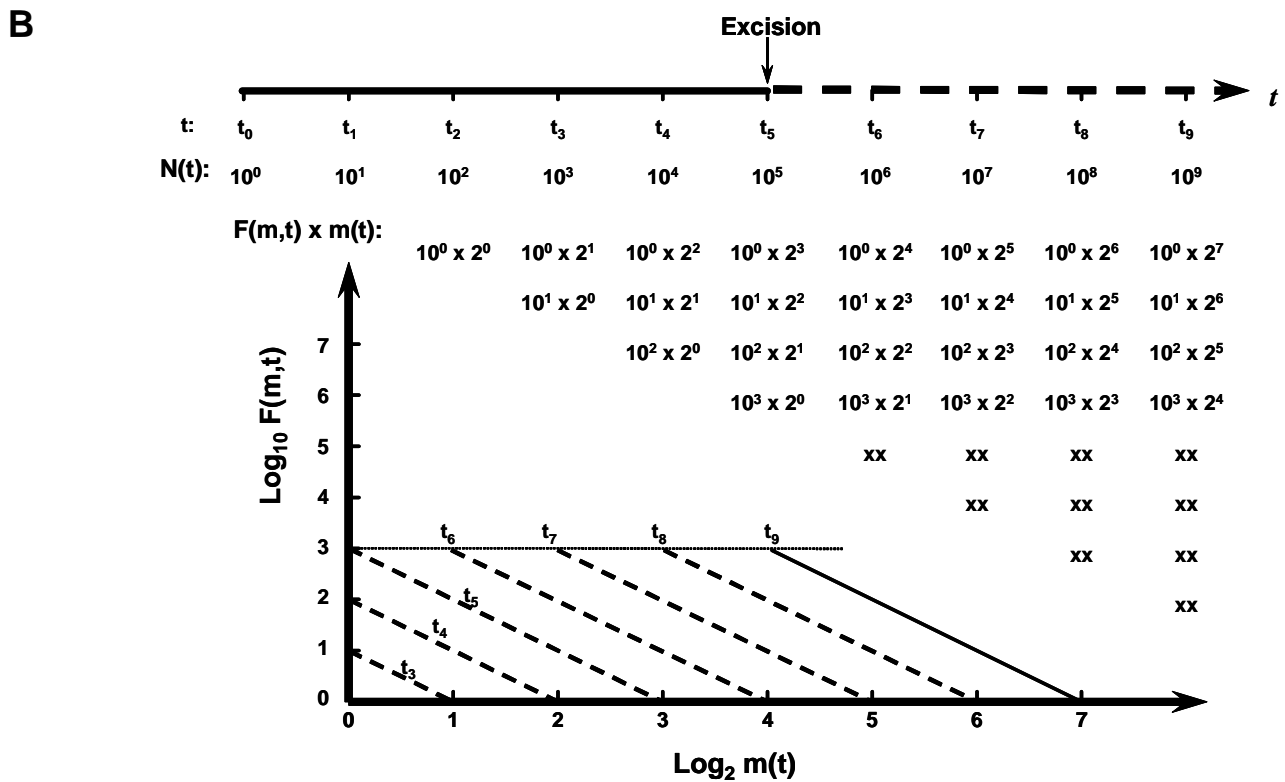

Fig. 2. Growth curves for primary tumors of two groups of mice:. *A*, a group of 119 mice implanted with 2 x 10⁵ MCaK cells, and *B*, a group of 150 mice implanted with 3 x 10⁵ MCaK cells. Either curve resembles exponential pattern for the *early* portion of its history (dark symbols), and the dashed curve is generated by fitting these selected data points with an exponential growth model (R=0.995 for *A*, and 0.991 for *B*). Data points for the larger volumes (open symbols), which were not only few in number but also contained massive necrosis, are not included for the curve fitting.

**A**

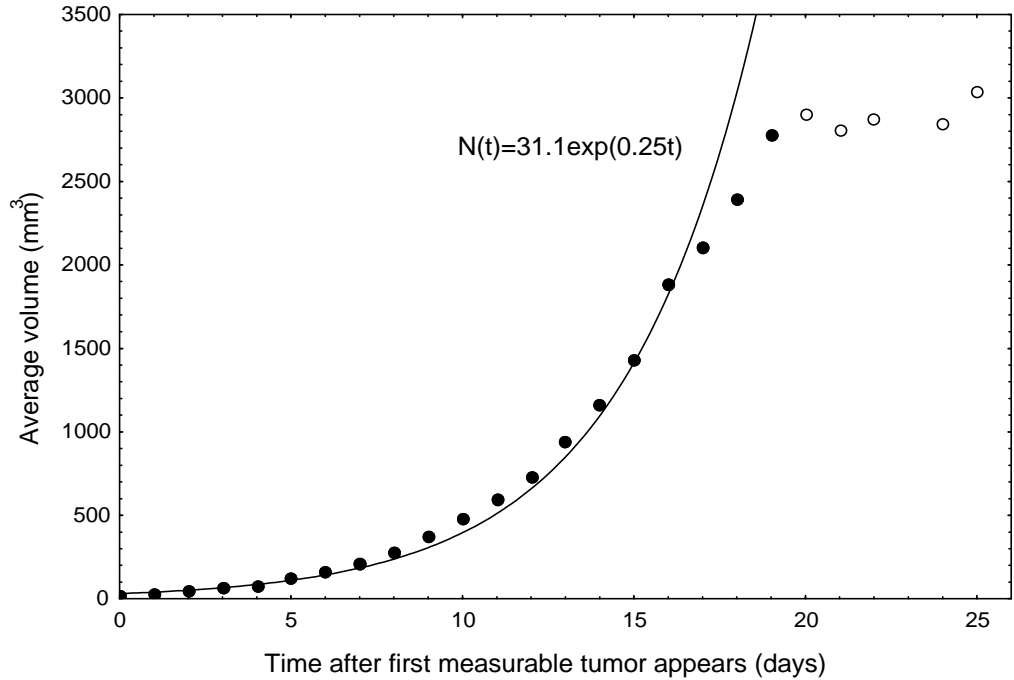

**B**

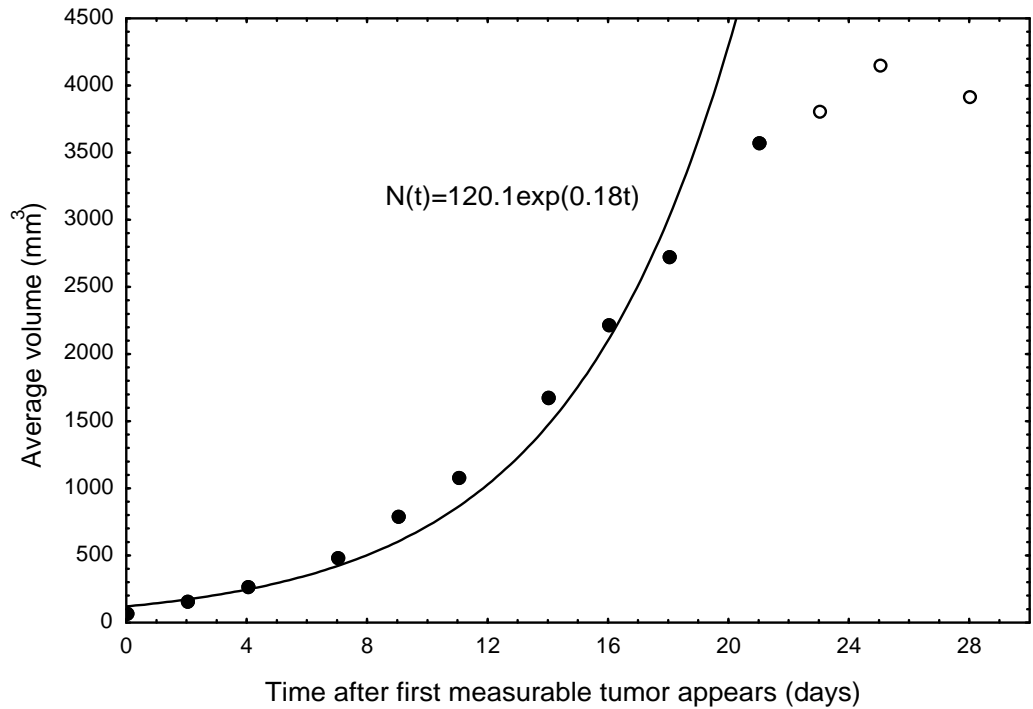

Fig. 3. *A*, semi-logarithmic display of the metastasis-free probability (*MFP*) for a group of mice implanted with MCaK cells, plotted as a function of their primary tumor size, *N*. The dotted curve is generated by fitting with the theoretical formula of Eq. (1) (R=0.934; quasi-Newton method). *B*, same data but plotted as log*MFP* vs. *N*. The solid curve is generated by fitting the data points with a linear-quadratic function based on Eq. (2) (R=0.934; quasi-Newton method), using also the numerical values for the linear term as obtained from *A* (dotted curve).

**A**

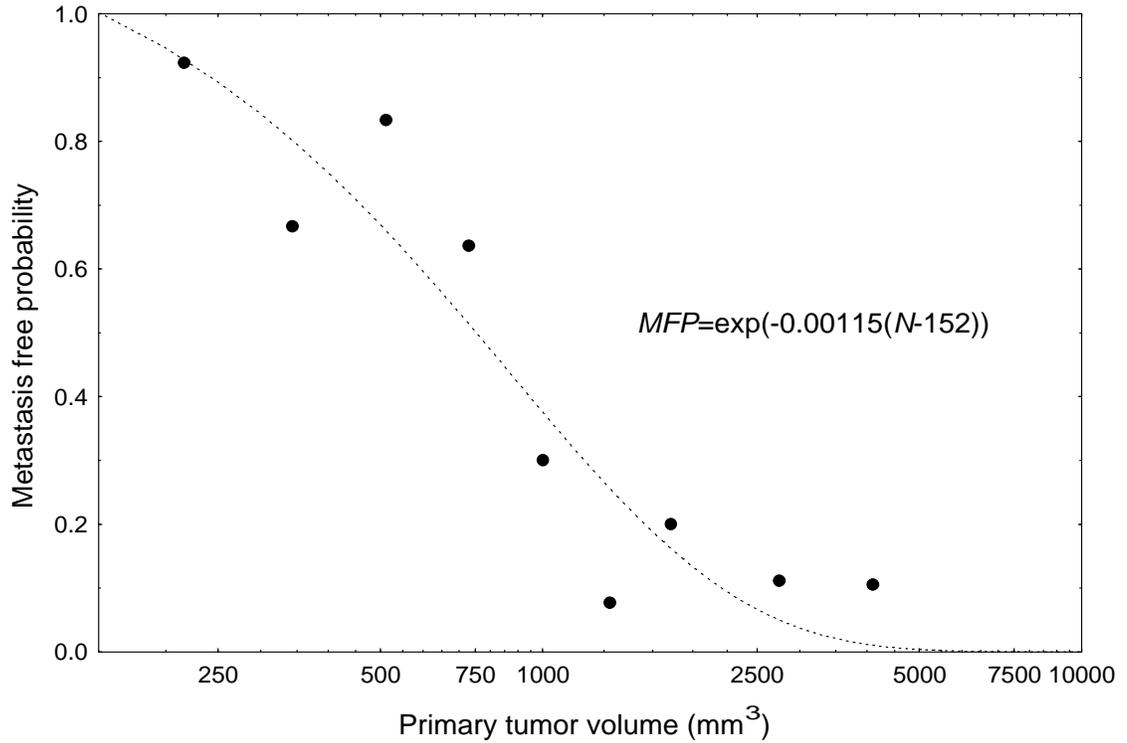

**B**

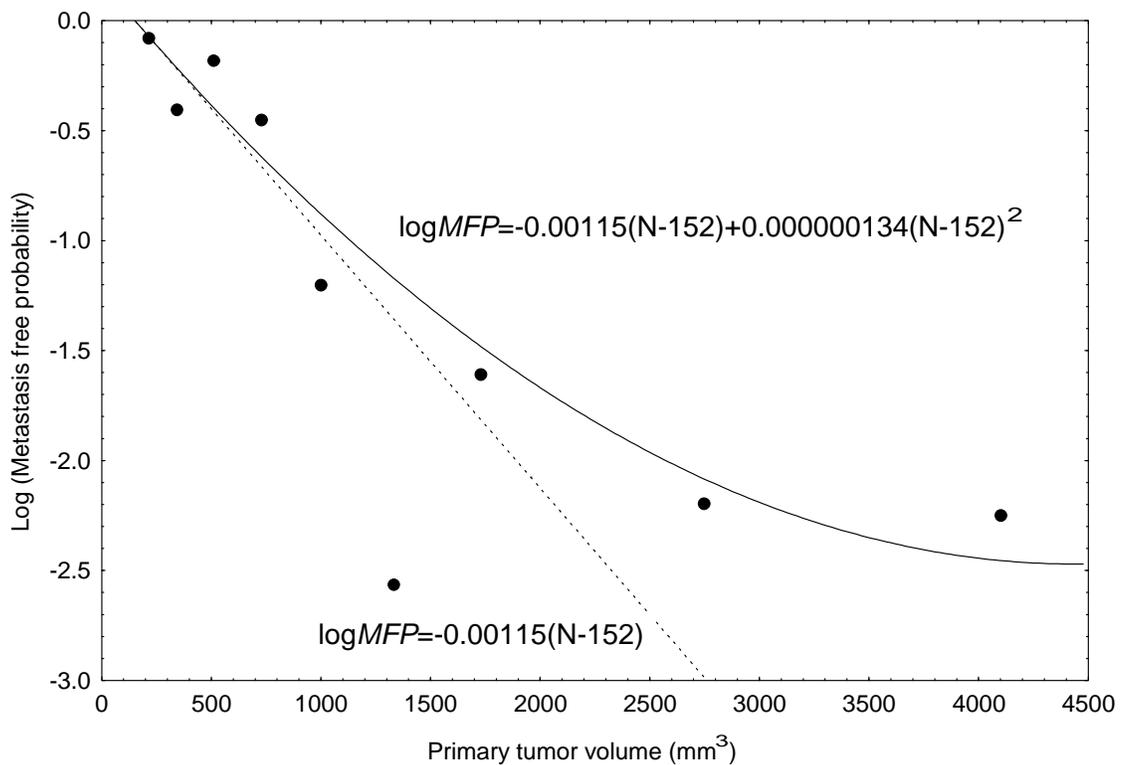

Fig. 4. Metastatic colony size distribution in the lungs of a mouse, whose flank was implanted with KHT tumor cells, and never underwent primary tumor excision. *A*, a semi-log display is shown, with the ordinate denoting the number of colonies counted, equivalent to the "frequency" function, $F(m,t)$. The data are fitted with an exponential distribution via non-linear regression (solid curve; R=0.906, quasi-Newton method). A parameter of 0.16 is found. The dashed curve corresponds to an exponential distribution with a parameter of 0.25, obtained retroactively from *B*. In *B*, logarithms of both the frequency function, $F(m,t)$, and the size density function, $R(m,t)$ are plotted against $\ln m$, then fitted with a linear function (quasi-Newton method: R=0.995 for $\ln R(m,t)$, and 0.923 for $\ln F(m,t)$), with the slope found to be -0.25 and -1.24, respectively.

**A**

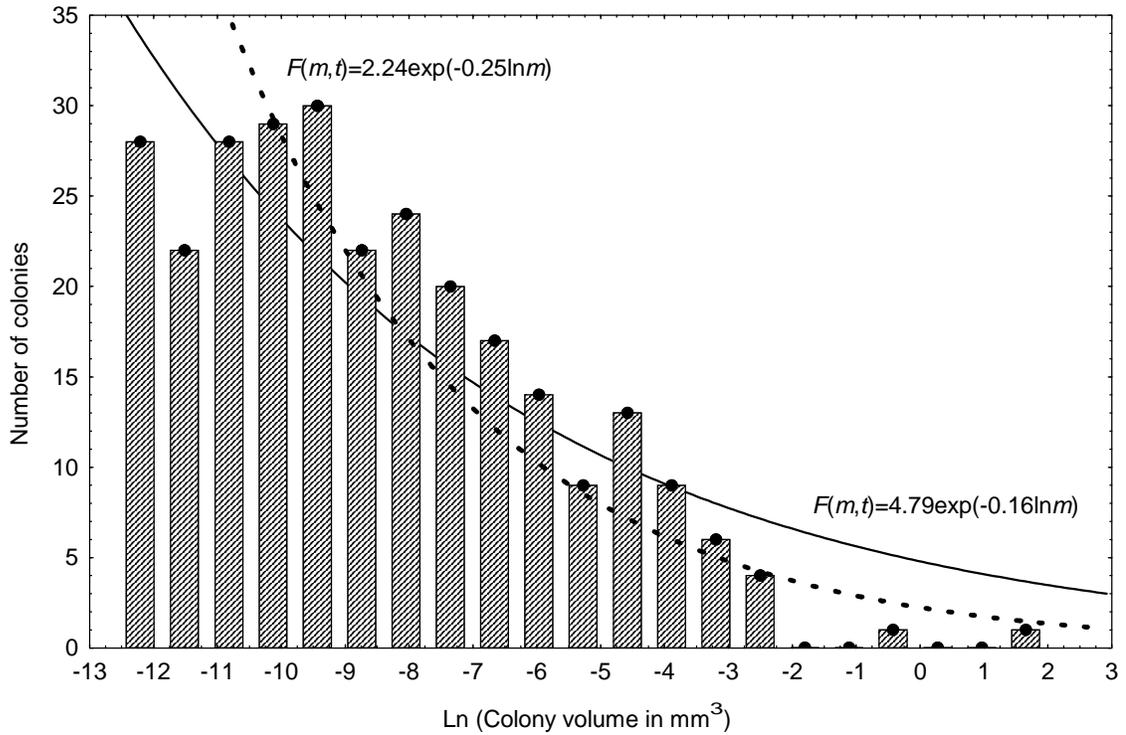

**B**

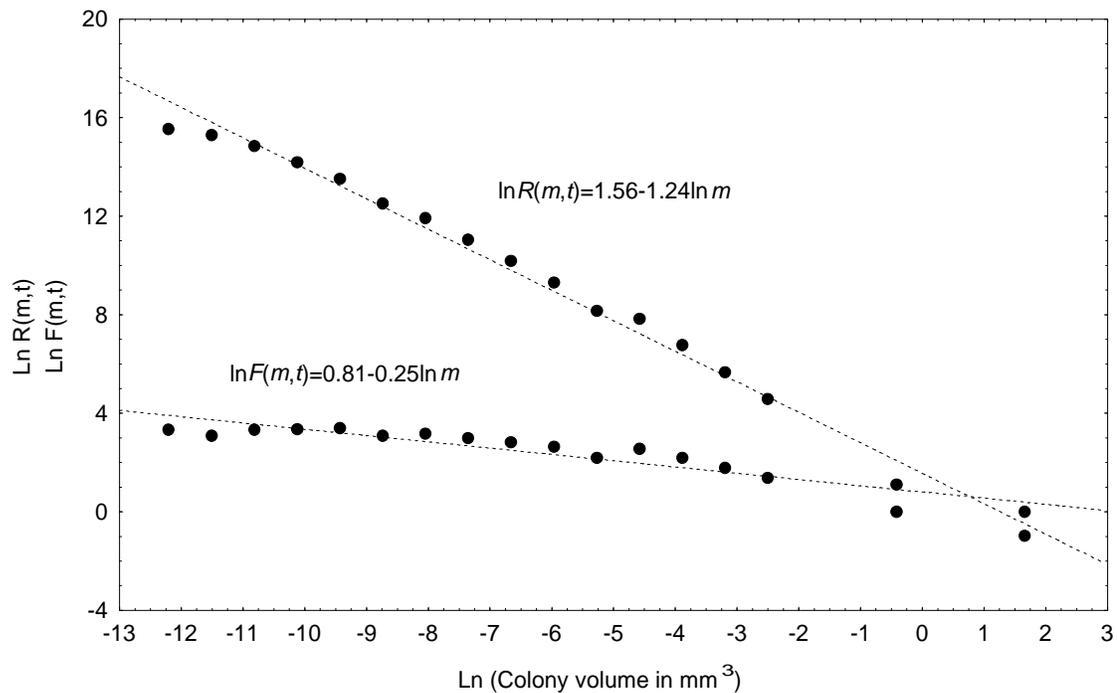

Fig. 5. Same data as in Fig. 4, showing only the size density function, $R(m,t)$. *A*, the curve labeled as residuals shows deviations of data points from the fitted theoretical curve. As can be inferred from Figs. 4A & 4B, more deviation at each extreme of the size spectrum are seen. *B*, fitting with a Gompertzian growth model for the primary tumor (R=0.998, Hooke-Jeeves pattern moves method). *C*, fitting with a logistic growth model for the primary tumor (R=0.998, Hooke-Jeeves pattern moves method).

**A**

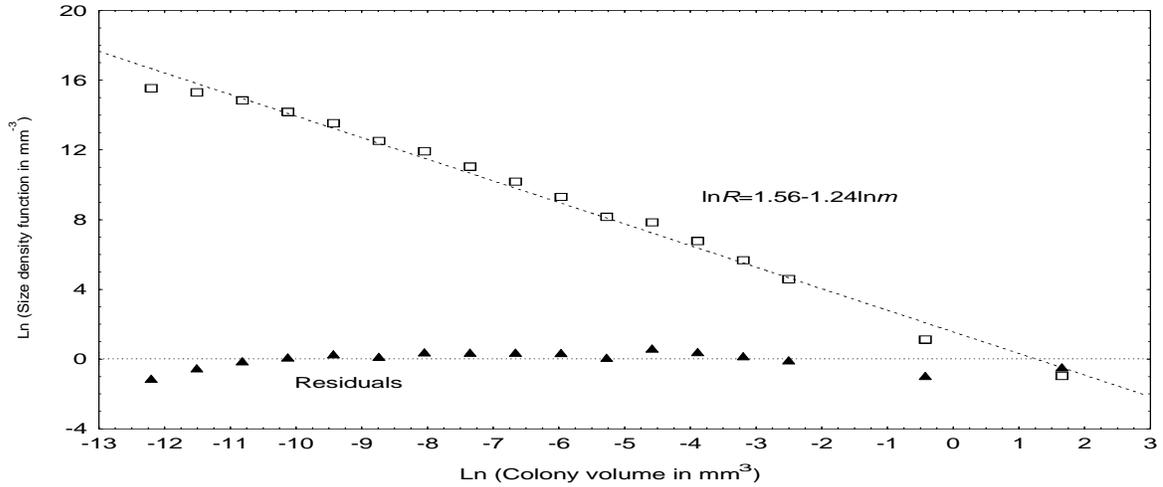

**B**

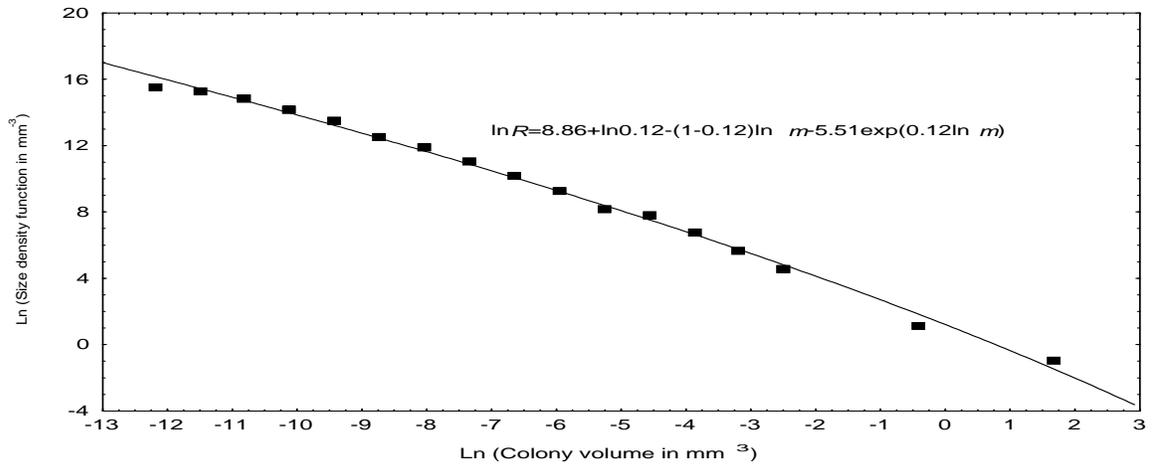

**C**

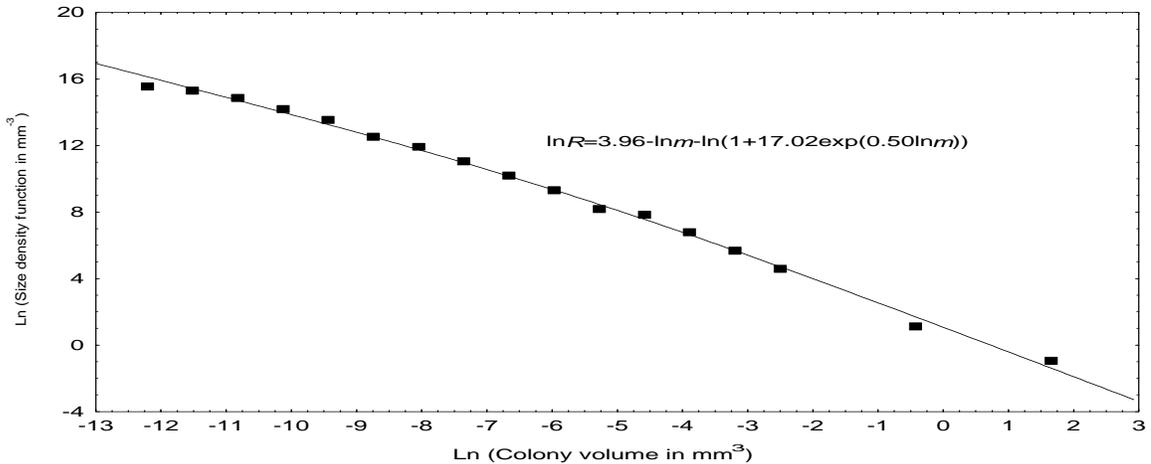

Fig. 6. Size density functions for two mice implanted with MCaK cells and excised of the primary tumor when the size reached 1 cm$^3$, and then sacrificed at 6 (circular symbols) and 12 (diamond symbols) days post-excision, respectively. Only data points over the larger volume range (solid symbols) are used to fit the theoretical straight line. Although a clear truncation of a straight line is not seen, as many smaller size colonies were observed (open symbols), there was absence of colonies smaller than about 4.5x10$^{-5}$ mm$^3$ (corresponding to -10 on the abscissa).

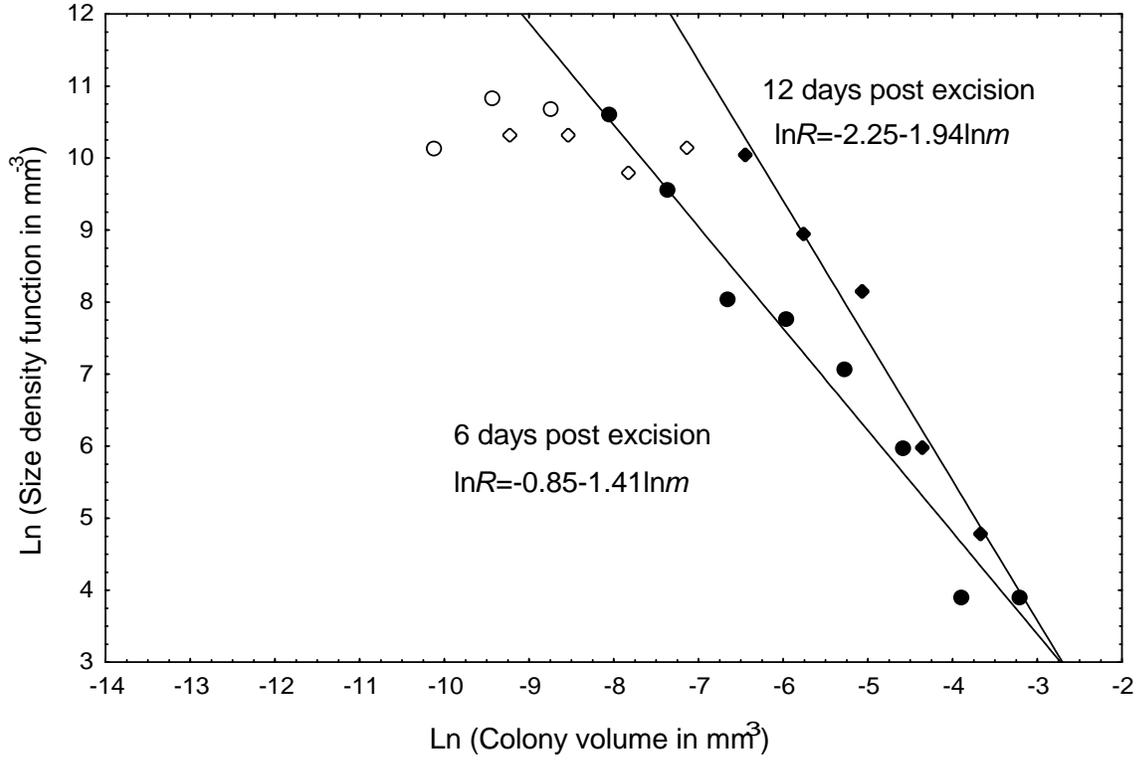